\documentstyle[12pt]{article}
\author{{\large M.Yu.Khlopov{\small $^{1,2,3,4}$}\thanks{e-mail: mkhlopov@orc.ru}
 ,~R.V.Konoplich{\small $^{1,2,3,4}$}\thanks{e-mail: konoplic@orc.ru}
 ,~S.G.Rubin{\small $^{2,4}$}\thanks{e-mail: sgrubin@orc.ru}
 ~and~A.S.Sakharov{\small $^{2,4}$}\thanks{e-mail: sakhas@landau.ac.ru}}}
\title{{\huge{\bf Formation of Black Holes in First Order Phase Transitions}}}
\date{{\small{\it $^1$Dipartimento di Fisica I Universita' di Roma "La 
Sapienza", P-le A.Moro,2,I-00185 Rome,Italy\\
$^2$Center for CosmoParticle Physics "Cosmion"\\ 
$^3$Institute of Applied Mathematics,
Miusskaya Pl.4, 125047 Moscow, Russia\\ 
$^4$Moscow Engineering Physics Institute 
(Technical University),  Kashirskoe Sh.31, 115409 Moscow, Russia}}}

\begin{document}
\maketitle

\begin{abstract}
A new mechanism of black hole formation in a first order phase transition is
proposed. In vacuum bubble collisions the interaction of bubble walls leads
to the formation of nontrivial vacuum configuration. The consequent collapse
of this vacuum configuration induces the black hole formation with high
probability. Observational constraints on the spectrum of primordial black
holes allow to obtain new nontrivial restrictions on parameters of inflation
models with first order phase transitions.
\end{abstract}

\section{Introduction}

At present time black holes (BH) can be created only by a gravitational
collapse of compact objects with mass more than about three Solar mass \cite
{1}. However at the early stage of evolution of the Universe there are no
limits on the mass of BH formed by several mechanisms. The simplest one is a
collapse of strongly inhomogeneous regions just after the end of inflation 
\cite{2}. Another possible source of BH could be a collapse of cosmic
strings \cite{3} that are produced in early phase transitions with symmetry
breaking. The collisions of the bubble walls \cite{4,5} created at phase
transitions of the first order can lead to a primordial black hole (PBH)
formation.

We discuss here new mechanism of PBH production in the collision of two
vacuum bubbles. The known opinion of the BH absence in such processes is
based on strict conservation of the original O(2,1) symmetry. Whereas
there are ways to break it . Firstly, the radiation of scalar waves
indicates the entropy increasing and hence the permanent breaking of the
symmetry during the bubble collision. Secondly, the vacuum decay due to
thermal fluctuation does not possess this symmetry from the beginning. The
simplest example of a theory with bubble creation is a scalar field theory
with two non degenerated vacuum states. Being stable at a classical level,
the false vacuum state decays due to quantum effects, leading to a
nucleation of the bubbles of true vacuum and their subsequent expansion \cite
{6}. The potential energy of the false vacuum is converted into a kinetic
energy of the bubble walls thus making them highly relativistic in a short
time. The bubble expands till it collides with another one. As it was shown
in \cite{4,5} a black hole may be created in the collision of several
bubbles. Our investigations show that BH can be created as well with a
probability of order unity in the collisions of only two bubbles. It
initiates the enormous production of BH that leads to essential cosmological
consequences discussed below.

In Section 2 the evolution of the field configuration in the collisions of
bubbles is discussed. The BH mass distribution is obtained in Section 3. In
Section 4 cosmological consequences of the BH production in bubble
collisions at the end of inflation are considered.

\section{Evolution of field configuration in collisions of true \qquad vacuum
bubbles}

Consider a theory where a probability of false vacuum decay equals $\Gamma $
and energy difference between the false and true vacuum equals $\rho _V$.
The vacuum decay proceeds through the nucleation of bubbles of new phase
separated from the false vacuum outside by initially unmoving walls. The
wall of the bubble increases quickly its velocity up to the speed of light $%
v=c=1$ due to conversion of the false vacuum energy into its kinetic one.

Let us discuss dynamics of collision of two true vacuum bubbles that have
been nucleated in points $({\bf r}_1,t_1),({\bf r}_2,t_2)$ and which are
expanding into false vacuum. Following papers \cite{4,7} let us assume for
simplicity that the horizon size is much greater than the distance between
the bubbles. Just after collision mutual penetration of the walls up to the
distance comparable with its width is accompanied by a significant potential
energy increase \cite{8}. Then the walls reflect and accelerate backwards.
The space between them is filled by the field in the false vacuum state
converting the kinetic energy of the wall back to the energy of the false
vacuum state and slowdown the velocity of the walls. Meanwhile the outer
area of the false vacuum is absorbed by the outer wall, which expands and
accelerates outwards. Evidently, there is an instant when the central region
of the false vacuum is separated. Let us note that the FVB does not possess
spherical symmetry at the moment of its separation from outer walls but wall
tension restores the symmetry during the first oscillation of FVB. As it was 
shown 
in \cite{7}, the further evolution of FVB consists of several stages:

1) FVB grows up to the definite size $D_M$ until the kinetic energy of its
wall become zero;

2) After this moment the false vacuum bag begins to shrink up to a minimal
size $D^{*}$;

3) Secondary oscillation of the false vacuum bag occurs.

The process of periodical expansions and contractions leads to energy losses
of FVB in the form of quanta of scalar field. It has been shown in the \cite
{7,9} that only several oscillations take place. On the other hand,
important note is that the secondary oscillations might occur only if the
minimal size of the FVB would be larger than its gravitational radius, $%
D^{*}>r_g$. The opposite case ($D^{*}<r_g$ ) leads to the BH creation with
the mass about the mass of the FVB. As we will show later the probability of
BH formation is almost unity in a wide range of parameters of theories with
first order phase transitions.

\section{Gravitational collapse of FVB and BH creation}

Consider in more details the conditions of converting FVB into BH. The mass $%
M$ of FVB can be calculated in a framework of a specific theory and can be
estimated in a coordinate system $K^{\prime }$ where the colliding bubbles
are nucleated simultaneously. The radius of each bubble $b^{\prime }$ in
this system equals to half of their initial coordinate distance at first
moment of collision. Apparently the maximum size $D_M$ of the FVB is of the
same order as the size of the bubble, since this is the only parameter of
necessary dimension on such a scale: $D_M=2b^{\prime }C$. The parameter $C\le 1$
is obtained by numerical calculations in the framework of each theory, but
its numerical value does not affect significantly conclusions.

One can find the mass of FVB that arises at the collision of two bubbles of
radius:

\begin{equation}
\label{one}M=\frac{4\pi }3\left( Cb^{\prime }\right) ^3\rho _V 
\end{equation}
This mass is contained in the shrinking area of false vacuum. Suppose for
estimations that the minimal size of FVB is of order wall width $\Delta $.
The BH is created if minimal size of FVB is smaller than its gravitational
radius. It means that at least at the condition

\begin{equation}
\label{two}\Delta <r_g=2GM 
\end{equation}
the FVB can be converted into BH (where G is the gravitational constant).

As an example consider a simple model with Lagrangian

\begin{equation}
\label{three}L=\frac 12\left( \partial _\mu \Phi \right) ^2-\frac \lambda
8\left( \Phi ^2-\Phi _0^2\right) ^2-\epsilon \Phi _0^3\left( \Phi +\Phi
_0\right) . 
\end{equation}
In the thin wall approximation the width of the bubble wall can be expressed
as $\Delta =2\left( \sqrt{\lambda }\Phi _0\right) ^{-1}$. Using (2) one can
easily derive that at least FVB with mass

\begin{equation}
\label{four}M>\frac 1{\sqrt{\lambda }\Phi _0G}
\end{equation}
should be converted into BH of mass M. The last condition is valid only in
case when FVB is completely contained in the cosmological horizon, namely $%
M_H>1/\sqrt{\lambda }\Phi _0G$ where the mass of the cosmological horizon at
the moment of phase transition is given by $M_H\cong m_{pl}^3/\Phi_{0}^2$. Thus 
for the potential (3) at
the condition $\lambda >(\Phi_0/m_pl)^2$ the BH is formed. This condition is 
valid for any realistic
set of parameters of theory. Let us find the mass and velocity distributions
of such BHs, supposing its mass is large enough to satisfy the inequality
(2). Apparently these distributions depend on coordinates and times of
nucleation of the bubbles. The probability of the collision of the
bubbles, which have been nucleated at the distance 
$\mid {\bf r_1}- {\bf r_1}\mid $ from each other has
the form: 
\begin{equation}
\label{five}
\begin{array}{c}
dP=dP_1\cdot dP_2\cdot P_{-}, \\ 
dP_1=\Gamma dt_1d^3r, \\ 
dP_2=\Gamma dt_24\pi |{\bf r}_2-{\bf r}_1|^2d|{\bf r}_2-{\bf r}_1|,
\end{array}
\end{equation}
where $dP_1$ is the probability of the bubble nucleation with coordinates $(%
{\bf r}_1,t_1)$, $dP_2$ is the probability of nucleation of the second
bubble at the distance $|{\bf r}_2-{\bf r}_1|\equiv 2b$ from first one (we
have integrated over angles assuming the space isotropy). The factor $%
P_{-}=\exp \left( {-\Gamma \Omega }\right) $ determines the probability,
which takes into account the absence of additional bubbles in 4- dimensional
region $\Omega $. In the following we consider the probability density of
the false vacuum decay being a free parameter. The region $\Omega $ will be
calculated below. Integrating(5) and assuming time independence of vacuum
decay probability we obtain

\begin{equation}
\label{six}dP/V=32\pi \Gamma ^2e^{-{\Gamma \Omega }}b^2dt_1dt_2db. 
\end{equation}

Here $V$ is the volume of cosmological horizon at the moment of the phase
transition. In the following we choose the $K^{\prime }$ system mentioned
above. The velocity of the system is $v=\left( t_1-t_2\right) /2b$ and
evidently $v$ is also the velocity of FVB (or BH). The radius of the colliding
bubbles is given by $b^{\prime }=b/\gamma $ , $\gamma =\left( 1-v^2\right)
^{-1/2}$. By using of (1) and (6), it is easy now to obtain the FVB
distribution in terms of new variables $M$, $v$ , $t$ , where $M$ is the
mass of FVB (or BH) created in the bubble collision, is its velocity and $t$ is
the first moment of the bubble contact: 
\begin{equation}
\label{seven}dP/VdvdM=\int \frac{64\pi }3\Gamma ^2e^{-\Gamma \Omega }\gamma
^4\left( \frac M{C\rho _v}\right) ^{1/3}\frac 1{C\rho _v}dt
\end{equation}
In order to determine the 4-dimensional area $\Omega $ we will use some
reasonable approximation. Namely, let us assume that every bubble that has
reached the sphere of radius $b^{\prime }$ with center in the point O at the
moment $t^{\prime }$ of first bubbles contact prevents the creation of FVB.
With this assumption the area $\Omega $ is determined as 
\begin{equation}
\label{eight}\Omega =\int_0^{t^{\prime }}d\tau ^{\prime }d^3{\bf r}^{\prime
}\theta \left( r^{\prime }+\tau ^{\prime }-b^{\prime }-t^{\prime }\right)
=\frac \pi 3\{\left( b^{\prime }+t^{\prime }\right) ^4-b^{\prime 4}\}
\end{equation}
Parameter $b'$ is related with the mass $M$ according to (1), the time $t$ in 
the $K'$
system equals to $\gamma t$. Therefore, after integration over time the 
distribution
of FVB in mass and velocity takes the form 
\begin{equation}
\label{nine}
\begin{array}{c}
dP/VdvdM=
\frac{64\pi }3\Gamma ^2\exp \left[ -\Gamma \frac \pi 3\left( \frac M{C\rho
_v}\right) ^{4/3}\right] \gamma ^4\left( \frac M{C\rho _v}\right)
^{1/3}\frac 1{C\rho _v}I, \\ I=\int_{t_{-}}^\infty d\tau \exp \{-\frac \pi
3\Gamma \left[ \left( \frac M{C\rho _v}\right) ^{1/3}+\gamma \tau \right]
^4\}, \\ 
t_{-}=\left( 1+v\right) \gamma \left( \frac M{C\rho _v}\right) ^{1/3}
\end{array}
\end{equation}

Let us compare a volume $V_{bag}$ containing one FVB and a volume $%
V_{bubble} $ of one bubble at the end of the phase transition. After
numerical integration of expression (9) we get 
\begin{equation}
\label{ten}V_{BH}\cong V_{bag}\cong 3.9\Gamma ^{-3/4}. 
\end{equation}
On the other hand, the average volume per one bubble is 
\begin{equation}
\label{11}V_{bubble}=\frac 43\pi \left( \frac 3\pi \right) ^{3/4}\Gamma
^{-3/4}\cong 4.0\Gamma ^{-3/4}, 
\end{equation}
where we assume the bubbles have a spherical form. The expected equality $%
V_{bag}$ = $V_{bubble}$ is fulfilled to conclude that our approximations are
correct. The distribution (9) can be rewritten in more convenient form in
terms of non dimensional mass $\mu \equiv \left( \frac \pi 3\Gamma \right)
^{1/4}\left( \frac M{C\rho _v}\right) ^{1/3}$:

\begin{equation}
\label{12}
\begin{array}{c}
\frac{dP}{\Gamma ^{-3/4}Vdvd\mu }=64\pi \left( \frac \pi 3\right) ^{1/4}\mu
^3e^{\mu ^4}\gamma ^3J(\mu ,v), \\ J(\mu ,v)=\int_{\tau _{}}^\infty d\tau
e^{-\tau ^4},\tau _{-}=\mu \left[ 1+\gamma ^2\left( 1+v\right) \right] .
\end{array}
\end{equation}
The numerical integration of (12) revealed that the distribution (9) 
has gaussian like shape with narrow maximum.
For example the number of BH with mass 30 times greater than 
the average one is
suppressed by factor $10^5$. Average value of the non dimensional mass is equal 
to $\mu=0.32$. It allows to relate the average mass of BH and volume containing 
the BH at the moment of the phase transition:

\begin{equation}
\label{MV}\left\langle M_{BH}\right\rangle =\frac C4\mu ^3\rho
_v\left\langle V_{BH}\right\rangle \simeq 0.012C\rho _v\left\langle
V_{BH}\right\rangle , 
\end{equation}
Remind that the constant $C$ is the model dependent value less then or order
unity.

\section{First order phase transitions in the early Universe}

Inflation models ended by a first order phase transition hold a dignified
position in the modern cosmology of early Universe (see for example \cite
{10,11}). The interest to these models is due to, that such models are able
to generate the observed large-scale voids as remnants of the primordial
bubbles for which the characteristic wavelengths are several tens of Mpc. 
\cite{11}. A detailed analysis of a first order phase transition in the
context of extended inflation can be found in \cite{12}. Hereafter we will
be interested only in a final stage of inflation when the phase transition
is completed. Remind that a first order phase transition is considered as
completed immediately after establishing of true vacuum percolation regime.
Such regime is established approximately when at least one bubble per unit
Hubble volume is nucleated. Accurate computation \cite{12} shows that first
order phase transition is successful if the following condition is valid .

\begin{equation}
\label{14}Q\equiv \frac{4\pi }9\left( \frac \Gamma {H^4}\right)
_{t_{end}}=1 
\end{equation}
Here $\Gamma$ is the bubble nucleation rate. In the framework of first order
inflation models the filling of all space by true vacuum takes place due to
bubble collisions, nucleated at the final moment of exponential expansion.
The collisions between such bubbles occur when they have comoving spatial
dimension less or equal to the effective Hubble horizon $H_{end}^{-1}$ at
the transition epoch. If we take $H_0=100hKm/\sec /Mpc$ in $\Omega =1$
Universe the comoving size of these bubbles is approximately $%
10^{-21}h^{-1}Mpc$. In the standard approach it believes that such bubbles
are rapidly thermalized without leaving a trace in the distribution of
matter and radiation. However, in the previous section it has been shown
that for any realistic parameters of theory, the collision between only two
bubble leads to BH creation with the probability closely to 100\% . The mass of 
this BH is given by (see (13))
\begin{equation}
\label{15}M_{BH}=\gamma _1M_{bub} 
\end{equation}
where $\gamma _1\leq 10^{-2}$ and $M_{bub}$ is the mass that could be
contained in the bubble volume at the epoch of collision in the condition of
a full thermalization of bubbles. The discovered mechanism leads to a new
direct possibility of PBH creation at the epoch of reheating in first order
inflation models. In standard picture PBHs are formed in the early Universe
if density perturbations are sufficiently large, and the probability of PBHs
formation from small post- inflation initial perturbations is suppressed
exponentially. Completely different situation takes place at final epoch of
first order inflation stage; namely collision between bubbles of Hubble size
in percolation regime leads to PBHs formation with masses

\begin{equation}
\label{16}M_0=\gamma _1M_{end}^{hor} 
\end{equation}
where $M_{end}^{hor}$ is the mass of Hubble horizon at the end of inflation.
According to (13) the initial mass fraction of this PBHs is given by
relationship:

\begin{equation}
\label{17}\beta _0=\gamma _1/e\approx 6\cdot 10^{-3} 
\end{equation}
A suppression $e^{-1}$ in comparison with (13) is included in order to avoid
a possibility of secondary bubbles nucleation inside FVB, which collapses
into PBH. The expression (16) can be rewritten in terms of inflation energy 
scale $%
H_{end}$ by the following manner:

\begin{equation}
\label{18}M_0=\frac{\gamma _1}2\frac{m_{pl}^2}{H_{end}} 
\end{equation}
For example, according to (18) and for typical value of upper limit of $%
H_{end}\approx 4\cdot 10^{-6}m_{pl}$ the initial mass fraction $\beta $ is
contained in PBHs with mass $M_0\approx 1g$.

On the radiation dominated stage the relative contribution of PBHs to total
cosmological density grows as a scale factor, and it means that at the moment

\begin{equation}
\label{19}t_1\approx \frac 1{\beta _0^2H_{end}} 
\end{equation}
over 50\% of matter are contained in PBHs. Since PBHs behave as dust-like
matter, and the state equation of the Universe has dust like form $p=0$ from the
moment $t_1$ . The PBHs dominated dust like stage ends at the moment of full
evaporation of PBHs at the moment:

\begin{equation}
\label{20}t_2\approx\frac 1{g_{*}}\left( \frac{M_0}{m_{pl}}\right) ^3t_{pl} 
\end{equation}
where $g_{*}$ is effective number of massless degrees of freedom at this
time \cite{13,14,2}. In the case of PBHs dominance they could form another
ones with greater masses. However the probability of more massive PBHs
formation would be negligible because of very small amplitude of initial
post inflation density perturbations $\delta \simeq 10^{-6}$.

There are a number of well-known limits, cowering various mass ranges on the
maximum allowed mass fraction of PBHs \cite{15,16,17,18,19,20,21}. Some are 
imposed at the
present epoch and some at earlier stages such as nucleosynthesis. These
constraints fall into two categories, those from the effect of Hawking
radiation and those from the gravitational effects. The evaporation of PBHs
via thermal emission has potentially observable astrophysical consequences.
Observations have placed limits on the maximum fraction of PBHs allowed at
evaporation. PBHs with mass $M_{ev}\leq 5\cdot 10^{14}g$ will have
evaporated before the present epoch. PBHs more massive that this will not
have experienced significant evaporation, and their present density should
not close the Universe ($\Omega _{PBH}<1$ ).

Ones generally suppose, that evaporation proceeds until the PBH vanishes
completely \cite{21}, but there are various arguments against this proposal 
\cite{22,23,24,25,26,27}. If one supposes that BH evaporation leaves a stable relic, it
normally is assumed to have a mass of order $m_{rel}=km_{pl}$ , where $%
k\simeq 1\div 10^3$. Let we derive the density of PBH relics from PBHs which
have been formed at the percolation epoch after first order inflation. Since
the probability of such PBHs formation is large, then there are quickly
coming on the domination regime, realized the early dust like stage, which
is ended at the moment of PBHs evaporation. The mass fraction of the
Universe going to relics becomes:
\begin{equation}
\label{21}\alpha _{rel}=k\frac{m_{pl}}{M_0}\sqrt{\frac{t_{eq}}{t_2}} 
\end{equation}
where $t_{eq}=3.2\cdot 10^{10}h^{-4}s$ is the moment when the densities of
matter and radiation become equal. It is clear that matter dominance can not
be turn on before $t_{eq}$. Consequently the inequality $\alpha _{rel}<1/2$
must be valid. By using (18), (20) and (21) we can express this condition in
the following form

\begin{equation}
\label{22}\frac{H_{end}}{m_{pl}}<\frac{0.4\gamma _1}{g_{*}^{1/5}k^{2/5}}%
\left( \frac{t_{pl}}{t_{eq}}\right) ^{1/5}\leq 7\cdot 10^{-12}h^{4/5}\frac{%
\gamma _1}{g_{*}^{1/5}k^{2/5}} 
\end{equation}
On the other hand the restriction \cite{15} on the relative contribution of PBH 
with mass $M<10^{11}$g. into cosmological density 
\begin{equation}
\label{23}\beta \left( M\right) <10^{-8}\left(\frac{10^{11}g}M\right) 
\end{equation}
implies that a significant fraction of the Universe can go to the PBHs with
the masses are not exceed $10^4g$. Thus, all entropy of our Universe could
be produced by such evaporating PBHs. Thereby the PBHs with masses $M<10^4g$
should be able to arise with probability of the order unit and without any
contradiction with observations. Last speculations imply the following
restriction on the first order inflation energy scale:

\begin{equation}
\label{24}\frac{H_{end}}{m_{pl}}\geq 10^{-9}\gamma _1 
\end{equation}
Two conditions (22) and (24) are incompatible, and if the relics hypothesis
is valid, there is a serious problem for all inflation models with reheating
by a first order phase transition.

\section{ Conclusion}

In this paper it was shown that BH creation takes place even in the
collision of two vacuum bubbles with high probability. The proposed
mechanism of BH formation puts severe restrictions on the scenario of the
evolution of the Universe. As an example, consider the class of cosmological
models, in which an inflation ends by a first order phase transition. The
horizon at the end of the inflation is about $H_{end}>4\cdot 10^{-6}m_{pl}$,
what may be estimated by COBE normalized reconstruction of inflation
potential. As it was evaluated above , produced black holes have then masses
about 1 g.

On the other hand there are arguments that black holes do not evaporate
completely leaving at the end of evaporation stable remnants with masses of
the order Planck mass.

Putting together these two conjectures one finds the serious trouble for the
considered class of inflation models. Evaporation of PBHs, formed in the
first order phase transition at the end of inflation, should results in
dramatic overproduction of stable remnants. Their contribution into the
present cosmological density should correspond to $10^{-14}g/cm^3$ , being
by 15 orders of magnitude above the observational upper limits. So one have
to conclude that the effect of the false vacuum bag mechanism of PBH
formation makes impossible the coexistence of stable remnants of PBH
evaporation with the first order phase transitions at the end of inflation.

\bigskip \bigskip
\noindent {\bf Acknowledgments} \par
\noindent This work was supported by part by the scientific and educational 
center
''Cosmion''. M.Yu.K. and R.V.K. are grateful to I Rome University and III
Rome University for hospitality and support.

\end{document}